\documentclass[useAMS,usenatbib]{mn2e}

\usepackage{graphicx}
\usepackage{amssymb}

\newcommand{\thisgrb}{GRB\,050716}
\newcommand{\swift}{\textit{Swift}}
\newcommand{\Tzero}{\ensuremath{\mathrm{T}_0}}
\newcommand{\Tnine}{\ensuremath{\mathrm{T}_{90}}}

\newcommand{\plusminus}[2]{\ensuremath{^{+#1}_{-#2}}}
\newcommand{\nh}{\ensuremath{N_\mathrm{H}}}
\newcommand{\ebv}{\ensuremath{E_{B-V}}}
\newcommand{\av}{\ensuremath{A_V}}
\newcommand{\nuc}{\ensuremath{\nu_{\mathrm{c}}}}
\newcommand{\num}{\ensuremath{\nu_{\mathrm{m}}}}
\newcommand{\alphao}{\ensuremath{\alpha_{\mathrm{o}}}}
\newcommand{\betao}{\ensuremath{\beta_{\mathrm{o}}}}
\newcommand{\chisqred}{\ensuremath{\chi^2_{\mathrm{red}}}}
\newcommand{\Ec}{\ensuremath{E_{\mathrm{c}}}}
\newcommand{\Ep}{\ensuremath{E_{\mathrm{p}}}}

\title[\thisgrb]{The early and late-time spectral and temporal evolution of \thisgrb}

\author[E. Rol et al.]
{E.~Rol,$^1$ 
J.~P.~Osborne,$^1$
K.~L.~Page,$^1$ 
K.~E.~McGowan,$^2$ 
A.~P.~Beardmore,$^1$
\newauthor
P.~T~.O'Brien,$^1$ 
A.~J.~Levan,$^3$ 
D.~Bersier,$^{4,5}$
C.~Guidorzi,$^{5,6}$
F.~Marshall,$^7$
\newauthor
A.~S.~Fruchter,$^4$
N.~R.~Tanvir,$^{1,3}$ 
A.~Monfardini,$^5$
A.~Gomboc,$^{5,8}$
\newauthor
S.~Barthelmy,$^7$ and
N.~P.~Bannister$^1$
\\
$^1$University of Leicester, University Road, Leicester LE1 7RH, UK \\
$^2$School of Physics \& astronomy, University of Southamption, Southampton, SO17 3BJ, UK \\
$^3$Department of Physics, Astronomy and Mathematics, University of Hertsfordshire, College Lane, Hatfield, Herts, AL9 10AB, UK  \\
$^4$Space Telescope Science Institute, 3700 San Martin Drive, Baltimore, MD 21218, USA \\
$^5$Astrophysics Research Institute, Liverpool John Moores University, Twelve Quays House, Birkenhead, CH41 1LD, UK \\
$^6$Osservatorio Astronomico di Brera, INAF, Via Brera 28, 20121, Milano, IT. \\
$^7$GSFC, USA \\
$^8$Faculty of Mathematics and Physics, University of Ljubljana, Jadranska 19, 1000 Ljubljana, SI.
}

\begin{document}

\date{}

\pagerange{\pageref{firstpage}--\pageref{lastpage}} \pubyear{}

\maketitle

\label{firstpage}

\begin{abstract}

  We report on a comprehensive  set of observations of Gamma Ray Burst
  050716,   detected  by  the   \swift\  satellite   and  subsequently
  followed-up rapidly in X-ray,  optical and near infra-red wavebands.
  The  prompt emission is  typical of  long-duration bursts,  with two
  peaks in a  time interval of $\Tnine = 68$ seconds  (15 -- 350 keV).
  The  prompt emission  continues at  lower flux  levels in  the X-ray
  band, where several smaller flares can be seen, on top of a decaying
  light~curve that exhibits an  apparent break around 220 seconds post
  trigger. This  temporal break is roughly coincident  with a spectral
  break.  The latter  can be related to the  extrapolated evolution of
  the  break  energy  in  the  prompt $\gamma$-ray  emission,  and  is
  possibly the manifestation  of the peak flux break  frequency of the
  internal  shock  passing  through  the  observing  band. A
    possible 3~$\sigma$ change in  the X-ray absorption column is also
    seen  during this  time.   The late-time  afterglow behaviour  is
  relatively standard,  with an electron  distribution power-law index
  of $p = 2$; there is no  noticable temporal break out to at least 10
  days.   The broad-band  optical/nIR  to X-ray  spectrum indicates  a
  redshift  of  $z \gtrsim  2$  for  this  burst, with  a  host-galaxy
  extinction  value of  $\ebv \approx  0.7$ that  prefers  an SMC-like
  extinction curve.

\end{abstract}


\section{Introduction}

One    of     the    main    goals    of     the    \swift\    mission
\citep{gehrels2004:apj611:1005}  is to  obtain  early-time information
for  gamma-ray   burst  (GRB)  afterglows  in   X-ray  and  optical/UV
wavelengths,  using its  rapid and  autonomous slewing  capability. In
addition,  longer term  monitoring, especially  in X-rays,  has become
more feasible than previously, with a dedicated facility like \swift.

Results of the early observations include the rapid decline of (X-ray)
afterglows      in      the       first      tens      of      minutes
\citep{tagliaferri2005:nature436,  goad2006:aa449:89}.  An explanation
for this  behaviour is prompt  emission seen from angles  further away
from our  line of sight  \citep{kumar2000:apj541}.  It is  likely that
emission from  the afterglow itself has  not risen enough  so early to
contribute significantly to the measured flux, although there are some
exceptions where  the (X-ray) flux  of the late-time  external forward
shock     is     already    visible     from     early    times     on
\citep[eg][]{obrien2006:apj647:1213}.   A  thorough understanding  of
this phenomenon requires detailed descriptions of the light~curve (and
broad-band spectral) behaviour early on, preferentially with as little
as possible  interference from other  phenomena such as  X-ray flares.
These   X-ray    flares   are   seen    in   a   number    of   bursts
\citep[eg][]{burrows2005:science309},    and    are   now    generally
interpreted   as    continuing   activity   of    the   inner   engine
(\citealt{king2005:apj630:113};  but  see also  \citealt{piro2005:apj623:314},
who  interpreted  the  X-ray  rebrightenings  as  the  result  of  the
beginning  of the  afterglow).  Late time  monitoring  allows the
construction of multiple  epochs of combined near infra-red/optical/UV
and  X-ray  broad-band spectra  (occasionally  including radio  data),
building a more complete picture of GRB afterglows than before.

We present here a full  analysis of \thisgrb. This includes the prompt
(burst)  emission in  gamma-rays and  early X-ray  data, for  which we
performed time-resolved  spectroscopy. We present the  late time X-ray
behaviour of the afterglow, as well as late time optical observations,
which we then combine to form a broad-band spectrum. The observations
and their analysis are described in Section \ref{section:obsana}, and in
Section   \ref{section:discussion}  we  discuss   the  results   of  the
observations. In Section \ref{section:summary}, we summarise our findings
and draw our conclusions.

In the following, $1\,\sigma$ errors are used except where
noted. The temporal and spectral power~law indices $\alpha$ and $\beta$
are defined by $F \propto t^{-\alpha} \nu^{-\beta}$, and the photon
index $\Gamma = 1+\beta$. All \swift\ data have been reduced using the
\swift\ software version 2.4 within the HEAsoft software (version
6.0.5), and the corresponding CALDB files.

\section{Observations and analysis} \label{section:obsana}

\thisgrb\  was detected  by the  \swift\ Burst  Alert  Telescope (BAT,
\citealt{barthelmy2005:ssrv120:143}) on  July 16, 2005,  at 12:36:04 UT.
All times mentioned  in this text are relative  to this \Tzero, except
where noted.  The spacecraft slewed and started observing in the X-ray
and optical/UV bands  96 and 99 seconds later,  respectively.  The XRT
\citep{burrows2005:ssrv120:165}  was  able  to  immediately  locate  a
position  for  the  X-ray  afterglow on-board,  allowing  ground-based
telescopes to perform rapid, deep follow-up observations.

\subsection{BAT analysis} \label{section:bat}

\begin{figure}
\includegraphics[width=\columnwidth]{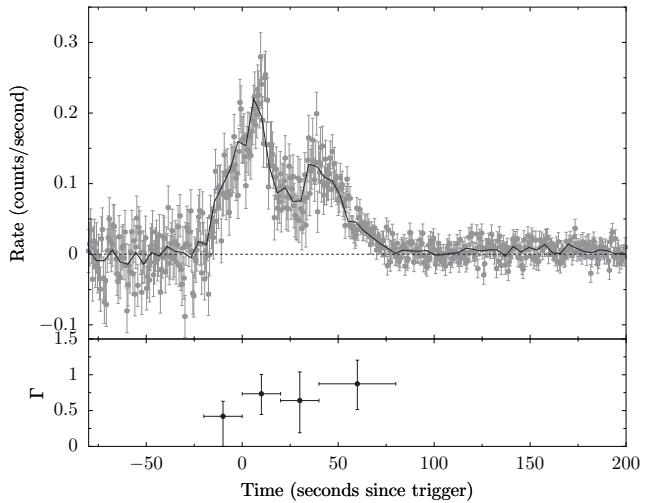}

\caption{\textbf{Top  panel}  The   BAT  15--350  keV  mask-weighted
light~curve, with  the two peaks around  9 and 39  seconds. Points are
0.5-second  binned  light~curve, while  the  overplotted  line is  the
4-second  binned  light~curve.  The  latter shows  some  evidence  for
possible emission after 80 seconds. \textbf{Bottom panel} The spectral
evolution of  the prompt emission in  the 15--150 keV  range; shown is
the photon index for a power~law model with an exponential cutoff.
\label{figure:batlc}
}

\end{figure}

BAT data were reduced using  the BAT software tools within the \swift\
software package.  A  PHA to PI energy conversion  was performed first
with the  \texttt{bateconvert} tool,  which ensures the  conversion is
quadratic, rather than linear.  We subsequently corrected the data for
hot  pixels. An  additional  systematic error  correction was  applied
before  spectral fitting.   The 15--350  keV light~curve  is  shown in
Figure \ref{figure:batlc}.

The  light~curve shows an  increase similar  to the  typical Fast-Rise
Exponential-Decay (FRED) behaviour seen  for many other bursts, with a
peak around 9  seconds after trigger and the start of  the rise of the
light~curve some 30 seconds earlier.  A notable difference is that the
rise  is not  as  steep as  usually seen  for  a FRED.   A similar  FRED
light~curve is  superposed on the tail  of the first  one, producing a
second peak 39  seconds after the trigger.  After 80  seconds, most of the
prompt  emission  has disappeared  in  the  15--350  keV energy  band.
\Tnine\ in this energy band for  this burst is $66 \pm 1$ seconds.
A 4-second binned light~curve  shows some emission still apparent past
80 seconds,  up to  200 seconds,  which is mostly  found in  the lower
energy bands.

We have  subdivided the  gamma-ray data into  4 temporal  sections and
obtained  spectral fits  for each  of  these. The  spectral shape  was
fitted  with a  power~law,  which  results in  an  acceptable fit.   A
power~law with an exponential cut-off, however, provides a better fit,
while  a fit with  a Band  model \citep{band1993:apj413:281}  does not
improve  the fit  compared  to either  one  of those  models. The  BAT
spectral energy  range and low number  of energy bins do  not allow an
accurate  determination of  the  cut-off energy  \Ec\  in the  cut-off
power~law model (for the first section,  it is in fact a lower limit),
but there  is an indication  that the spectrum  softens, with
\Ec\ gradually becoming lower,  and the photon index $\Gamma$ changing
from about  0.4 to  0.9 during the  prompt gamma-ray emission,  over a
period of $\approx 50$ seconds.

To  verify this  spectral evolution,  we  have fitted  the later  time
sections  (C  \&  D)  with   the  model  parameters  (other  than  the
normalisation) fixed at those obtained from the earlier sections (A \&
B).  By  then freeing the parameters  and applying an  F-test, we find
that  there is  indeed a  change in  the spectrum  at a  99.9\% level.
Performing a  similar test to  the first and  last sections (A  \& D),
with the results obtained from the fit to the total section (-20 -- 80
s), also indicates a changing spectrum at a 99.9\% confidence level.

Such  softening behaviour  of the  prompt spectrum  has  been commonly
found before in other bursts \citep{ford1995:apj439:307}.  The results
are  detailed in Table  \ref{table:gammaray-spectra}, and  include the
result  of   a  spectral   fit  to  the   whole  time   span.   Figure
\ref{figure:batlc} shows the spectral evolution of the prompt emission
between 15 and 150 keV as well.

\begin{table}
\caption{Results of spectral fits to four sections of the prompt
gamma-ray emission, as well as that of the combined fit.
\label{table:gammaray-spectra}
}
\begin{tabular}{lrlll}
Section &
Time &
$\Gamma$ &
cutoff energy &
\chisqred 
\\
 &
(sec) &
 &
(keV) & 
\\
\hline

A &
-20 -- 0 &
0.42 \plusminus{0.21}{0.60} &
221  ($> 6.7$) &
1.05
\\

B &
0 -- 20 &
0.74 \plusminus{0.27}{0.29} &
125 \plusminus{178}{49} &
0.86 
\\

C &
20 -- 40 &
0.64 \plusminus{0.40}{0.45} &
105 \plusminus{305}{48} &
0.86 
\\

D &
40 -- 80 &
0.87 \plusminus{0.33}{0.36} &
65.6 \plusminus{47.5}{20.9} &
0.95 
\\

A -- D &
-20 -- 80 &
0.85 \plusminus{0.21}{0.23} &
150 \plusminus{169}{54} &
0.76
\\
\hline

\end{tabular}
\end{table}

\subsection{XRT analysis} \label{section:xrt}

XRT data were reduced from level 1 to level 2 with the \swift\
software task \texttt{xrtpipeline}. Data obtained in Windowed Timing
(WT) mode and Photon Counting (PC) mode have been used in all our
following analysis. Counts were grouped by 20 per  bin for the spectra.
WT mode  data for the  light curve have  been grouped by 40  counts per
bin, and  PC mode light curve data  have been grouped by  20 counts per
bin.

We  have  subdivided  our  analysis  into an  early-time  part  and  a
late-time part. We designate the early-time part as the section of the
X-ray afterglow  that shows flares  (probably related to  inner engine
activity, see  also the discussion  below), while the  late-time part
shows a  relatively smooth decay and  is presumed to  be the afterglow
emission from a forward shock. The  data for the early part extends up
to almost 1ks, with most of  the data having been obtained in WT mode.
The late-time part  starts some 4ks after the  trigger and extends out
to almost three weeks.

\subsection{Early XRT data} \label{section:earlyxrt}

WT mode  observations started at 103  seconds, and extended  up to 516
seconds. After this, the count-rate  was low enough for the instrument
to automatically switch to PC  mode. WT mode data were extracted using
a rectangular region with a length of 93\arcsec\ centred on the source
and  along  the  readout  direction,  with  an  equally  sized  region
off-source serving as the  background determination. PC mode data have
been  extracted with  a 47\arcsec\  circular aperture  instead  of the
default 71\arcsec\,  because of  a contaminating close-by  source. The
first few  100 seconds  of PC mode  data are  piled-up and we  used an
annular  extraction  region, with  a  12\arcsec\  inner  radius and  a
47\arcsec\ outer  radius.  In addition,  some bad columns  are located
near the source centre.  We used  default grades for WT mode (grades 0
to 2),  while we used only  grade 0 for  the first few 100  seconds of
PC-mode observations.

To determine  the combined correction factor for  the annular aperture
and bad  column loss, we  modelled the PSF using  \textsf{ximage}, and
calculated the ratio of  the integrated response between an unmodified
PSF and  that masked  out by  an inner 12\arcsec\  circle and  the bad
columns. In the same manner, we correct for the expected loss of using
only a 47\arcsec\ circular aperture. The combined correction factor is
4.47.

\subsubsection{Spectral evolution}

Similar to the BAT observations, we  find evidence for a change in the
spectral slope over time.  Modelling  the spectrum with a simple power
law, we  find that the  X-ray spectrum starts  with $\Gamma =  0.9 \pm
0.1$ (0.3 -- 10 keV), and the  slope then evolves to $\Gamma = 2.0 \pm
0.1$, with a change between 250 and 400 seconds post trigger.

If a break is expected in the X-ray band, a broken power~law model for
the spectra around this time is preferred. Fits with broken power~law
models, however, do not show a significant improvement in the fit, and
we have therefore used the results from the single power~law fits above.

Because there is still some flux in the BAT light~curve coincident
with the first 100 seconds or so of WT mode data, we have performed a
joined fit to the BAT and XRT data between 100 and 200 seconds post
trigger. A single power law fit has a \chisqred of $1.98$ (degrees of
freedom (DOF) = 61), but a Band model does a much better job
($\chisqred = 0.90$, DOF = 59), and results in $E_0 =
12.2\plusminus{5.4}{2.4}$~keV, or a peak energy of $\Ep =
13.1\plusminus{5.8}{2.6}$~keV.  The spectral indices are $\alpha =
-0.93\plusminus{0.08}{0.13}$ and $\beta =
-2.56\plusminus{0.37}{1.51}$.

Because  of the  low number  of counts  in the  BAT data,  the  fit is
dominated by  the XRT  data and the  outcome should be  taken somewhat
cautiously.

\subsubsection{Light curve} \label{sect:light-curve}

Since  there is a  change in  the spectral  slope halfway  through the
light~curve, we  cannot simply fit the count-rate  light~curve: a flux
calibrated light~curve  has to be  used.  To derive the  count-rate to
flux conversion factors, we have  used the spectra between 103 and 223
seconds   and  between   263   and  513   seconds   (see  also   Table
\ref{table:nh}), which provide the conversion values for the first and
last part of the light~curve,  respectively. For the middle section we
have used  the average of these  values: the rapid  change in spectrum
does not allow for a good  spectral fit. In all our following fits, we
have  modelled  the  two   most  pronounced  flares  with  two  simple
Gaussians. We also left out the last data point of the light~curve: it
is likely that  this point is the onset  of the later-time light~curve
(see  Section \ref{section:latexrt}),  or possibly  even the  start of
another flare.

A fit to the flux light curve with a single power~law results in
$\chisqred = 2.06$ (DOF = 123), while a broken power~law fit gives
$\chisqred = 1.83$ (DOF = 121).  Neither fit is good, but the broken
power~law does provide a significant improvement. A fit with a
smoothly broken power~law (either with the sharpness of the break
fixed or free) hardly improves this ($\chisqred = 1.78$ for
  both, DOF = 121 and 120, respectively), and an exponential fit is
worse ($\chisqred = 2.07$, DOF = 123).

The broken power~law fit results  in decay parameters $\alpha_1 = 0.91
\plusminus{0.30}{0.09}$ and  $\alpha_2 = 3.78 \plusminus{0.19}{0.15}$.
The temporal  break is found at $220  \plusminus{27}{4}$ seconds, just
before the  time of the spectral break.   This coincidence strengthens
the suggestion that  a broken power~law is a  reasonable model for the
underlying  light~curve. See Section  \ref{section:early} for  a more
detailed exploration of this break.

\begin{figure}
\includegraphics[width=\columnwidth]{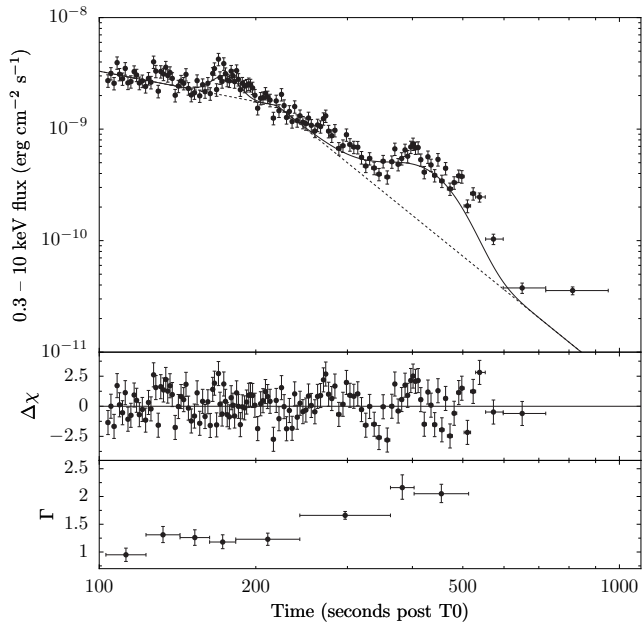}

\caption{ \textbf{Top  panel} 0.3--10 keV  early-time light~curve. The
last three points  are PC mode data, the  previous data points consist
of WT mode data.  A broken
power~law fit with two Gaussians is shown as the continuous line, with
its break time around 200  seconds post trigger (the Gaussians account
for the  two main flares, to  obtain a better overall  fit; the broken
power~law without  Gaussians is shown  as the dashed line).   The last
data  point  has  not been  fit,  since  this  likely belongs  to  the
later-time   light~curve. \textbf{Middle  panel}  The   fit  residuals,
expressed  in sigma  deviation. Even  with  the Gaussian  fits to  two
flares, the overall light  curve shows significant deviations from the
fit. \textbf{Bottom panel} The  evolution of the photon index $\Gamma$
of the  X-ray spectrum.
\label{figure:earlylc}
}

\end{figure}

The  light~curve  and  the  spectral  evolution are  shown  in  Figure
\ref{figure:earlylc}.   As commonly seen  in other  early X-ray
light~curves
\citep[eg][]{burrows2005:science309,nousek2006:apj642:389,obrien2006:apj647:1213},
this one exhibits  a few flares, which most  likely indicates activity
of  the burst  engine  itself.   This would  agree  with the  possible
emission  seen  in the  BAT  15--350 keV  range  between  100 and  200
seconds. Also visible is the  deviation of the underlying decay from a
power~law.

\subsubsection{X-ray absorption change?}

\begin{figure}
\includegraphics[angle=-90,width=\columnwidth]{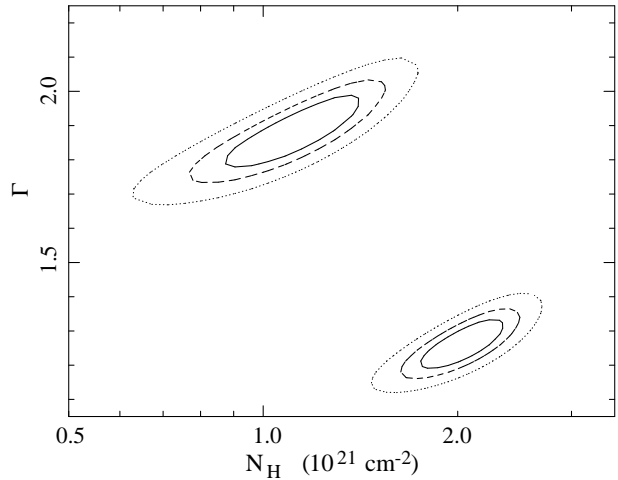}

\caption{Contour  plots for  the  1-$\sigma$, 90\%  and 99\%  combined
confidence  intervals for  the  photon index  $\Gamma$  and \nh.   The
contours  upper-left  are for  the  WT-mode  data  after the  observed
spectral  break (303-513  seconds),  the contours  in the  lower-right
corner  are pre-break  (103-253  seconds).  The  middle part  (253-303
seconds) is left out, since here  the change in spectral break is most
pronounced,  indicating  $\Gamma$   is  highly  variable.   Note  that
$\Gamma$  for the  first  data  section is  slightly  higher than  the
previously quoted $0.9$,  which is for the very  first part of WT-mode
data only.
\label{figure:contour}
}

\end{figure}

For the combined WT-mode data, we find no evidence for excess \nh\
above that of the estimated Galactic column of $1.1 \times 10^{21}
\mathrm{cm}^{-2}$ \citep{dickey1990:araa28:215}. Subdividing the WT-mode
into a pre- and a post- spectral break section indicates, however, a
modest amount of excess absorption at early times, while no excess
absorption is measured past the break: $\nh =
2.0\,(\pm\,0.2)\,\cdot\,10^{21}\,\mathbf{\mathrm{cm}^{-2}}$ and $\nh =
1.1 (\pm 0.2) \cdot 10^{21} \mathbf{\mathrm{cm}^{-2}}$ for the total
\nh\ at the two respective epochs.  While the observed change in
photon index could correlate with this change in \nh, a contour plot
(Figure \ref{figure:contour}) indicates it is likely that the column
density did change.  Something similar has been seen, for example, in
the early (prompt) X-ray emission for GRB\,000528
\citep{frontera2004:apj614:301}, or claimed for GRB\,050730
\citep{starling2005:aa442:21}.  An F-test for a model with \nh\ fixed
at the later time value (see also Section \ref{section:latexrt}),
compared to a model with \nh\ free to fit, gives a probability of
6.7e-5 for the two models to be consistent.  See the second part of
Table \ref{table:nh} for more details.

\begin{table*}
\caption{Results of spectral fits for the XRT WT-mode data. The second half
of the table details the results on variation of the column
density. See also Figure \ref{figure:contour}. A simple power~law was
used, with the Teubingen-Boulder model
\citep[][\texttt{tbabs}]{wilms2000:apj542:914} for the absorption.
\label{table:nh}
}
\begin{tabular}{llll}
\multicolumn{1}{c}{time}  & 
\multicolumn{1}{c}{$\Gamma$} & 
\multicolumn{1}{c}{\nh} & 
\multicolumn{1}{c}{$\chi^2_{\mathrm{red}}$} \\
\multicolumn{1}{c}{(sec)} &
 &
\multicolumn{1}{c}{($10^{21}$ cm$\mathbf{^{-2}}$)} &
 \\

\hline
103 -- 123  &  
$0.95 \pm 0.12$  &
$1.37 \plusminus{0.73}{0.66}$  &
0.509 (12.2/24)  \\

123 -- 143  &
$1.31 \plusminus{0.15}{0.14}$  &
$1.87 \plusminus{0.74}{0.67}$  &
0.899 (15.3/17)  \\

143 -- 163  &  
$1.26 \pm 0.14$  &
$1.18 \plusminus{0.66}{0.61}$  &
0.518 (9.84/19)  \\

163 -- 183  &  
$1.18 \plusminus{0.13}{0.12}$  &
$1.11 \plusminus{0.61}{0.57}$  &
0.698 (16.8/24)  \\

183 -- 243  &  
$1.23 \pm{0.11}$  &
$1.65 \plusminus{0.42}{0.39}$  &
0.801 (25.6/32)  \\

243 -- 363  &  
$1.66 \pm{0.07}$  &
$1.51 \plusminus{0.25}{0.23}$  &
1.09 (63.3/58)  \\

363 -- 403  &  
$2.16 \plusminus{0.23}{0.21}$  &
$1.88 \plusminus{0.56}{0.53}$  &
1.12 (13.5/12)  \\

403 -- 513  &  
$2.05 \plusminus{0.17}{0.16}$  &
$1.09 \plusminus{0.39}{0.36}$  &
1.09 (22.9/21)  \\

\hline
\multicolumn{4}{l}{\hspace{-0.5cm}\nh\ variation} \\

103 -- 223  &  
$1.26 \pm 0.05$  &
$2.04 \plusminus{0.21}{0.20}$  &
0.890 (111.2/125) \\

263 -- 513  &
$1.86 \plusminus{0.08}{0.07}$  &
$1.13 \plusminus{0.19}{0.18}$  &
1.053 (73.69/70)  \\

\hline

\end{tabular}
\end{table*}

\subsection{Late time XRT observations}
\label{section:latexrt}

\begin{figure}
\includegraphics[width=\columnwidth]{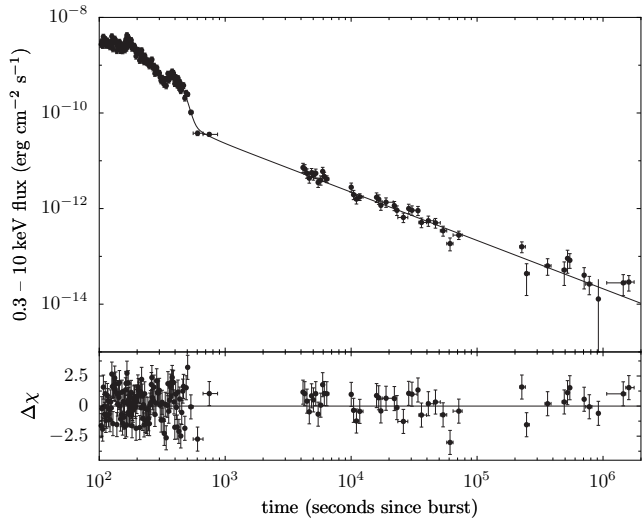}

\caption{\label{figure:xrtlc}
The complete X-ray light~curve between
0.3 and 10 keV. The continuous line is modelled using a broken
power~law and 2 Gaussians for the early time light~curve (see Section
\ref{section:earlyxrt}), and a single power~law added to that for the
late-time light~curve. The latter has a decay index of $0.99 \pm
0.02$.
\textbf{Bottom panel} The residuals of the fit,
expressed in sigma deviation.
}
\end{figure}

The XRT  data after 4ks are  not piled-up, and were  extracted using a
circular aperture  of 47\arcsec  radius and the  default grades  (0 to
12).   The  resulting 0.3\,--\,10  keV  light~curve  follows a  smooth
power-law decline,  with a  power-law decay index  of 0.99.   The full
light~curve  is shown  in Figure  \ref{figure:xrtlc} and  includes the
early time light~curve  as well.  There is no  indication of a shallow
part  ($\alpha \lesssim  0.5$) in  the light~curve  between  the steep
early decay and  the later time decay, as often  seen around this time
in the light~curve  \citep[eg][]{nousek2006:apj642:389}; this may have
been  missed if  it was  relatively short  and between  the  early and
late-time part.   There has been, however,  no need to  include such a
section  in  our  fits to  obtain  good  results.   There is  also  no
indication of a late-time break  before the light~curve is lost in the
background noise,  up to  $10^6$ seconds. A  spectral fit to  the data
between 10ks and 100ks, with \nh\ fixed at its Galactic value, results
in a good fit ($\chi^2_{\mathrm{red}} = 0.54$), with a photon index of
$\Gamma = 2.01 \plusminus{0.11}{0.10}$.

\subsection{Optical observations}
\label{section:optical}

Optical  observations were  automatically performed  with  the \swift\
Ultra       Violet       and       Optical      Telescope       (UVOT,
\citealt{roming2005:ssrv120:95})  in the  $UBVW1\,M2\,W2$  filters and
without  a filter.   The  prompt dissemination  of  the position  also
triggered  the Faulkes Telescope  North (FTN),  which began  the first
observations     of      its     automatic     follow-up     procedure
\citep{guidorzi2006:pasp188:288}   243  seconds   after   the  \swift\
trigger.  Finally,  near infra-red observations at  the United Kingdom
Infrared  Telescope (UKIRT) started  56 minutes  post trigger,  in the
$JHK$ bands,  and Gemini NIRI  $K$-band observations were  obtained 22
hours after \Tzero.  The log of our observations can be found in Table
\ref{table:optical-log}.

A faint,  fading source was detected  in the UKIRT  data, just outside
the  90\%  XRT  error  circle,  and identified  as  the  IR  afterglow
\citep{tanvir2005:gcn3632}.  The IR data  also gives the best position
for   the  GRB   and   its  afterglow.    This  is   RA~=~22:34:20.73,
Dec~=~+38:41:03.6 (J2000), with an estimated error of 0.4\arcsec. This
position  is derived  from the  NIRI $K$-band  image,  calibrated with
respect to the 2MASS survey \citep{skrutskie2006:aj131:1163}.

A single power~law fit to the three $K$-band points results in a
power~law decay index of $\alphao = 0.80\plusminus{0.13}{0.07}$, with
$\chi^2_{\mathrm{red}} = 0.689$. While different from the X-ray decay
slope at these times, the paucity of data points does not allow a
strong quantification of this difference. A fit with $\alphao$ fixed
at $\alpha_X = 0.99$, however, results in $\chi^2_{\mathrm{red}} =
2.58$.

We calculated the $JHK$ spectral index by shifting the data to a
common epoch using the previously calculated decay index, and
correcting for the estimated Galactic extinction
\citep{schlegel1998:apj500}. Conversion to flux was performed using
the Vega fluxes from \citet{fukugita1995:pasp107} for the optical and
from \citet{tokunaga2005:pasp117} for the nIR filters.  The fit shows
the spectrum to be inconsistent with a power-law spectrum ($\betao =
2.9$, but $\chisqred = 4.2$), indicative of intrinsic reddening. We
address this in Section \ref{section:broadband}.

\begin{table*}
\caption{\label{table:optical-log}   The   results   of  our   optical
follow-up campaign, covering the  wavelength range from near infra-red
to the near  ultraviolet. Other data collected through  the GCN network
provide only  upper limits similar to  those of the FTN  and UVOT, and
have therefore not been repeated here.}
\begin{tabular}{cccccc}
   Start date         &   days since burst  &  exposure time  &  filter  &
  magnitude    &  telescope/instrument  \\
(days UT, July 2005)  &      (mid-time)     &    (seconds)    &          & 
              &             \\
\hline
16.529991 &   0.01954   &  400    &    $i'$   &  $ > 21.5 $       &  FTN         \\
16.527721 &   0.00306   &  30     &    $R$    &  $ > 20.3 $       &  FTN         \\
16.527721 &   0.02239   &  540    &    $R$    &  $ > 22.3 $       &  FTN         \\
16.529911 &   0.02236   &  520    &    $B$    &  $ > 23.0 $       &  FTN         \\
16.526169 &   0.4315    &  5799   &    $V$    &  $ > 21.2 $       &  \swift/UVOT  \\
16.527859 &   0.4042    &  5507   &    $B$    &  $ > 22.1 $       &  \swift/UVOT  \\
16.527697 &   0.3988    &  5477   &    $U$    &  $ > 21.6 $       &  \swift/UVOT  \\
16.527534 &   0.4416    &  5053   &   $W1$    &  $ > 21.5 $       &  \swift/UVOT  \\
16.527373 &   0.4374    &  5486   &   $M2$    &  $ > 21.8 $       &  \swift/UVOT  \\
16.528044 &   0.4084    &  4383   &   $W2$    &  $ > 21.8 $       &  \swift/UVOT  \\
19.055751 &   2.7699    &  9480   &  'white'  &  $ > 22.5 $       &  \swift/UVOT  \\
16.579109 &   0.05594   &  300    &    $J$    &  $20.76 \pm 0.08$ &  UKIRT/UFTI  \\
16.581678 &   0.05937   &  450    &    $H$    &  $19.31 \pm 0.05$ &  UKIRT/UFTI  \\
16.564178 &   0.04272   &  600    &    $K$    &  $17.74 \pm 0.15$ &  UKIRT/UFTI  \\
16.583727 &   0.06055   &  300    &    $K$    &  $18.22 \pm 0.15$ &  UKIRT/UFTI  \\
17.421337 &   0.9275    & 3600    &    $K$    &  $20.5 \pm 0.2$   &  Gemini/NIRI \\
\hline
\end{tabular}
\end{table*}

\section{Discussion} \label{section:discussion}

\begin{table*}
\caption{\label{table:summary}
  Summary of the main results obtained in Section \ref{section:obsana}. 
  We have converted the spectral results from $\Gamma$ to $\beta$.
}
\begin{tabular}{llll}
\multicolumn{1}{c}{data section}  & 
\multicolumn{1}{c}{$\beta$} & 
\multicolumn{1}{c}{$\alpha$} & 
 \\
\hline

X-ray, 103 -- 223 seconds & 
0.26 $\pm$ 0.05 &
0.91 \plusminus{0.30}{0.09} 
\\

X-ray, 263 -- 513 seconds &
0.86 \plusminus{0.08}{0.07} &
3.78 \plusminus{0.19}{0.15} 
\\

X-ray, post 1 hour &
1.01 \plusminus{0.11}{0.10} &
0.99 $\pm$ 0.02 
\\

optical, post 1 hour &
2.94$^1$ &
0.80 \plusminus{0.13}{0.07} 
\\
\hline
$^1$ $\chisqred = 4.2$, therefore no error has been calculated.
\end{tabular}
\end{table*}

Here we discuss  the results obtained above. A  short summary of the
results is given in Table \ref{table:summary}.

\subsection{Early emission}
\label{section:early}

\thisgrb\ shows  characteristics as seen in  previous (\swift) bursts,
both   in   its   prompt    emission   and   the   (early)   afterglow
\citep{nousek2006:apj642:389,obrien2006:apj647:1213}.    The  flares,
combined  with the  multiple peaks  in the  gamma-ray  emission, imply
several  shocks,  likely   indicative  of  prolonged  engine  activity
\citep{king2005:apj630:113}.  The  early X-ray data  show a
power-law decline, that appears to be unrelated to the later afterglow
emission and  which is often explained as  prompt emission originating
from higher latitudes  at the emission source,  after the emission at
smaller angles has  faded away.

In the context of this model, we can test the curvature effect $\alpha
= 2  + \beta$  \citep{kumar2000:apj541}. We used  the values  pre- and
post-break  in  the XRT  data,  showing  that  by this  relation,  the
observed  $\alpha_1$   is  too  shallow  and   $\alpha_2$  too  steep.
\citet{zhang2006:apj642:354}  list several  possible  causes why  this
relation would  not hold. Most important  are the effect  of using the
trigger time  instead of the time  of the last  emission peak (thereby
assigning an  incorrect zeropoint  for a power  law fit), and  that of
superposition of tails of several emission peaks. Both causes could be
valid here, since the last  significant emission peaks some 40 seconds
after  the  trigger. However,  applying  such  a  correction for  both
effects would effectively produce  an ever shallower decay, making the
discrepancy larger. Possible causes for $\alpha$ being too shallow are
the    fact   that   one    is   looking    at   a    structured   jet
\citep{rossi2002:mnras332:945} or when the observed (X-ray) wavelength
regime is still below the cooling break. In the latter case, the decay
is given by $\alpha  \sim 1 + 3 \beta/2$ \citep{zhang2006:apj642:354}.
Only the very  early section of the X-rays  agrees with this relation,
with    $\alpha    =   0.91$    and    $\beta    =   -0.1$    (Section
\ref{sect:light-curve},  Table \ref{table:nh}).   Post-break, however,
the discrepancy  is even larger  in this scenario.   Potentially, both
the peak frequency and the cooling break have passed the X-ray band at
this  time. While  this  could explain  the  relatively long  spectral
transition, there is no direct  evidence for this.  The discrepancy at
later times, for either relation, might more likely be attributed to
an increasing influence of  the external shock (which likely dominates
the last  data), and the  last flare, which  hamper a good fit  to the
steeper tail of the early light curve.

Evolution of the early emission in the X-ray to gamma-ray energy range
has  been studied before,  eg by  \citet{frontera2000:apjs127:59}, who
looked at  the spectral  evolution for eight  \textit{Beppo}SAX bursts
which had been detected by both the Gamma-Ray Burst Monitor and one of
the Wide-Field Cameras. While their findings are similar, an important
advantage here is that we have  an early X-ray light curve which, with
provisions for some  flaring, can be fit with a  broken power law; the
break time  of this  power law is  roughly coincident with  the moment
\Ep\ passes through the X-ray  band. The WFC data generally cover only
the very  early part of the  burst, and any  underlying (broken) power
law is  not visibile.  As a  result, the \Ep\ evolution  could only be
deduced from the spectral changes for the \textit{Beppo}SAX data.

Variable X-ray absorption at early times in the X-ray spectra of GRBs
has been noted before in other GRBs
\citep{frontera2000:apjs127:59,amati2000:sci290:953,intzand2001:apj559:710,starling2005:aa442:21}.
\citet{lazzati2002:mnras330:383} have suggested that ionisation
strongly modifies the absorption properties of the surrounding
material. For this effect to be noticable, the material should be in a
compact region surrounding the GRB ($< 5$ parsec), and have an initial
column density $\nh > 10^{21} \mathrm{cm}^{-2}$. The change detected
here is $0.8 \cdot 10^{21} \mathrm{cm}^{-2}$ at redshift 0.  For a
redshift of $z = 2$ (see below), the column density would be $\nh
\approx 1.5 \cdot 10^{22} \mathrm{cm}^{-2}$ (at Solar meticallicity),
large enough to fulfill the $\nh > 10^{21} \mathrm{cm}^{-2}$
criterion.

\subsection{The late-time light~curve and broad-band spectrum}
\label{section:broadband}

From the combination of the X-ray temporal and spectral index around
5ks, we can infer a electron power~law index of $p = 2.0$ within the
standard fireball model
\citep[eg][]{meszaros1997:apj476:232,sari1999:apj519:17,chevalier1999:apj520:29}.
This assumes that 1) the afterglow is in the slow-cooling regime
($\num < \nuc$), 2) \nuc\ is below the X-rays and 3) the jet-break has
not manifested itself at this time. Other scenarios are easily ruled
out by invalidating relations between the two indices (see eg Table 1
in \citealt{zhang2004:ijmp19:2385}).  There is still a degeneracy
between the type of circumburst environment (a constant density medium
or a medium with a $R^{-2}$ dependency ($R$ the distance from the
centre) such as a stellar wind), which can potentially be resolved by
looking at the temporal and spectral indices below \nuc; in our case
the only candidate for that is the optical band, which is generally
found to be below \nuc\ at these times. To verify this is indeed the
case, we compared the X-ray and optical decay indices.

There is a 1.5$\sigma$ difference between the afterglow decay slope in
optical and X-rays, which is only marginally significant and does
allow \nuc\ to be below optical wavelengths. However, a cooling break
between the optical and X-ray waveband makes the picture
self-consistent: for an ISM-like, the decay index in optical should be
0.25 lower than that in X-ray, 0.74, which is certainly compatible
with $\alphao$ obtained from our fit to the $K$-band data. On the
other hand, for a wind-like medium, the optical decay index would have
to be 0.25 higher than that in X-ray, 1.24, and is therefore
incompatible with \alphao.

The non power-law behaviour of the near infra-red (nIR) spectrum
indicates reddening, which we estimate by examining the broad-band
spectral behaviour, by combining the X-ray, optical and nIR data.  For
this, we have extrapolated the optical and X-ray data to a common
epoch: we choose 0.04 days (3.5ks) after trigger, which is central to
the optical FTN and nIR UKIRT data, and lies just at the beginning of
the late-time X-ray data. We obtained an unabsorbed X-ray flux from
the data between 4ks and and 12ks (orbits 2 and 3 combined), which
were then extrapolated back to 0.04 days using the power~law decay
index of $0.99$.  The optical and nIR data were corrected for Galactic
extinction and converted to fluxes as before.  We then extrapolated to
0.04 days post trigger using our estimate for the optical power~law
decay index.

\begin{figure}
\includegraphics[width=\columnwidth]{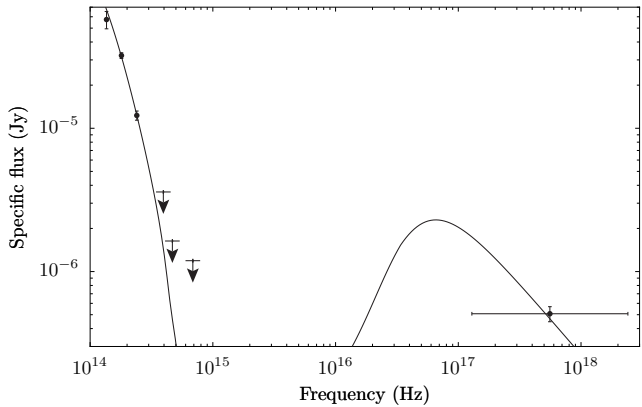}

\caption{The X-ray to optical/nIR broad-band spectrum of GRB\,050716, 0.04 days post burst. The best fit is shown for a broken power~law connecting the two
  wavelength ranges, with host-galaxy absorption for an
  SMC-like extinction curve at $z = 2$. The X-ray spectrum has been
  represented relatively simply by a single data point. This
    has the purpose of enhancing the \chisqred\ sensitivity on the
    nIR data instead of on the X-ray data, since we try to calculate
    the estimated optical/UV host-galaxy extinction.
\label{fig:bbspec}
}

\end{figure}

With  the scaled  fluxes, we  constructed the  broad-band  spectrum of
\thisgrb\ at 0.04 days (Figure \ref{fig:bbspec}). We then tried to fit
the  data with models  available within  the standard  fireball model,
allowing for intrinsic (UV/optical)  extinction within the host galaxy
and  a Lyman  cutoff  resulting from  its  redshift. No  spectroscopic
redshift  is known,  however.  While  in principal  one  can obtain  a
photometric redshift, there are too  many free parameters with the few
available data  to do this  accurately. We have therefore  obtained an
estimate of  the intrinsic extinction  for an afterglow  positioned at
various redshifts  (1, 2, 3,  \ldots, 7). We have  further constrained
the power~law slope connecting the  X-ray and optical/nIR regime to be
equal to that obtained by the  X-ray spectral fit alone, $\beta = 1.0$.

A single power~law, absorbed by host-galaxy extinction or
intergalactic Lyman extinction, is not able to reproduce the observed
broad-band spectrum. A spectral break between optical and X-ray is
required, as already found from the fits to the optical and X-ray
light~curves.  With the assumption that this break is indeed the
cooling break (with $\Delta \beta = 0.5$), we have kept the lower end
of this break fixed at a power~law index of $\beta_{\mathrm{opt}} =
0.5$, and the higher end at $\beta_{\mathrm{X}} = 1.0$).  Reasonable
fits are then obtained for redshifts above $z = 2$, with host-galaxy
extinction that follows an SMC-like extinction curve (\ebv\ = 0.5 --
0.7, \av\ = 1.5 -- 2.1, with lower values for higher redshifts. See
Figure \ref{fig:bbspec}). This estimate for the host-galaxy is higher
than on average found for GRB host galaxies, but not unknown
\citep[eg][]{levan2006:apj647:471}.  However, localised extinction or
a line-of-sight through an edge-on galaxy could easily explain the
somewhat larger than average host extinction.

The reduced $\chi^2$ is in all cases $\lesssim 1$ (for redshifts above
4, it becomes much less than 1), resulting in a lower limit on the
redshift of $z \gtrsim 2$.  A Galactic-type extinction curve
results in a steeper nIR slope than measured here, and
  provides a worse fit. The same gradual nIR slope also implies that
host-galaxy extinction should be present, and cannot be explained by a
Lyman break alone.  We further looked at the resultant extinction
value for small changes ($\pm 0.1$) in our assumed spectral index,
which results in changes of about 0.1 magnitude in \ebv (the value for
\chisqred, however, also increases considerably doing so). Finally,
from our best fit we find that the \nuc\ cannot be located directly
below the X-ray frequencies, but is restricted to $\sim 10^{16}$ Hz.

An \ebv\ of $0.7$ with an SMC-like extinction curve would imply an
\nh\ of $3.2 \cdot 10^{22}$ \citep{martin1989:aa215:219} in the host
galaxy.  Correcting for an assumed SMC-metallicity of $1/4$ Solar and
adjusting for the redshift of $z = 2$, this turns out to be not
measurable above the estimated Galactic extinction in our data, and
indeed, no excess extinction is measured past 300 seconds.

We can obtain a lower limit on the fluence and burst energy from the
BAT spectrum, which results in $5.9 \cdot 10^6$ erg cm$^{-2}$ and $5.6
\cdot 10^{52}$ erg, respectively. Both values are in the 15 -- 150 keV
energy range in the observers frame.  With a lower limit of $z \gtrsim
2$, the relation found by Amati
\citep{amati2002:aa390:81,amati2006:mnras372:233} gives $\Ep
\gtrsim 76$ keV in the observers frame, while we find $\Ep \gtrsim
170$ keV from the spectral fit (Table \ref{table:gammaray-spectra},
after conversion of the cut-off energy to \Ep).  Note that the latter
uses a power law with an exponential cut-off, not the Band function.
Using the previous fit values with a Band function, fixing $\beta =
-2$, and extrapolating outside the BAT spectral range (0.001 --
$10^{4}$ keV), we obtain an estimate of the isotropic energy of
$E_{\mathrm{iso}} \approx 2.3 \cdot 10^{53} \mathrm{erg}$, which is
almost four times larger than the 15 -- 150 keV estimate.  Using the
Amati relation, this would put the peak energy at $\sim 150$ keV in
the observer frame, assuming $z~\approx~2$.  This redshift estimate
therefore agrees reasonably well with the results expected from the
Amati relation, and indicates $z~\approx~2$ is a likely good estimate
for the redshift of this burst.

\section{Summary} \label{section:summary}

We have presented here a comprehensive analysis of the data available
on GRB\,050716.  The most remarkable feature of the early emission is
the spectral change, clearly visible in the X-rays, which can possibly
be traced back to the gamma-ray emission.  The temporal break
coincident with the spectral break points to the peak-flux frequency
passing through the X-rays at this time.  In addition, the X-ray
absorption before the break appears to be higher than after.  While
this could be an artificial result related to the break, care has been
taken to eliminate instrumental effects and the result appears to be
significant, possibly indicating ionisiation of surrounding
  medium.

At later times, our multi-wavelength data of \thisgrb\ shows a fairly
standard afterglow: the X-ray light curve and spectrum indicate $p =
2$, with no evidence for a break related to the broadening of the jet
outflow.  The optical data are rather limited, partly because the
optical afterglow is relatively faint: host-galaxy extinction with
\av\ = 1.5 -- 2.1 can account for the obtained
upper limits.
Though no  spectroscopic redshift has  been obtained for  \thisgrb, we
obtain a  lower limit of  $z \gtrsim 2$, which  appears to  be in
agreement with the Amati relation.

\section*{Acknowledgements}

We thank the referee for a careful reading of the manuscript, which
improved it overall.
ER, JPO, KLP, APB, AJL and NRT acknowledge financial support from
PPARC. The  authors like  to thank Simon  Vaughan for  additional help
with  parts of  the  data analysis.  This  work is  also supported  at
Pennsylvania State  University (PSU)  by NASA contract  NAS5-00136 and
NASA grant  NNG05GF43G, and at  the Osservatorio Astronomico  di Brera
(OAB) by  funding from  ASI on grant  number I/R/093/04.   "The United
Kingdom Infrared  Telescope is operated by the  Joint Astronomy Centre
on  behalf  of  the  U.K.   Particle Physics  and  Astronomy  Research
Council."  This  publication makes use  of data products from  the Two
Micron All Sky  Survey, which is a joint project  of the University of
Massachusetts    and   the    Infrared    Processing   and    Analysis
Center/California  Institute  of Technology,  funded  by the  National
Aeronautics  and   Space  Administration  and   the  National  Science
Foundation. We gratefully appreciate  the contributions of all members
of the \swift\ team.

\bibliographystyle{mn2e}
\bibliography{references}

\begin{thebibliography}{38}
\expandafter\ifx\csname natexlab\endcsname\relax\def\natexlab#1{#1}\fi

\bibitem[{{Amati}(2006)}]{amati2006:mnras372:233}
{Amati}, L., 2006.
\newblock \mnras, 372, 233

\bibitem[{{Amati} et~al.(2000)}]{amati2000:sci290:953}
{Amati}, L., et~al., 2000.
\newblock Science, 290, 953

\bibitem[{{Amati} et~al.(2002)}]{amati2002:aa390:81}
{Amati}, L., et~al., 2002.
\newblock \aap, 390, 81

\bibitem[{{Band} et~al.(1993)}]{band1993:apj413:281}
{Band}, D., et~al., 1993.
\newblock \apj, 413, 281

\bibitem[{{Barthelmy} et~al.(2005)}]{barthelmy2005:ssrv120:143}
{Barthelmy}, S.~D., et~al., 2005.
\newblock Space Science Reviews, 120, 143

\bibitem[{{Burrows} et~al.(2005{\natexlab{a}})}]{burrows2005:science309}
{Burrows}, D.~N., et~al., 2005{\natexlab{a}}.
\newblock Science, 309, 1833

\bibitem[{{Burrows} et~al.(2005{\natexlab{b}})}]{burrows2005:ssrv120:165}
{Burrows}, D.~N., et~al., 2005{\natexlab{b}}.
\newblock Space Science Reviews, 120, 165

\bibitem[{{Chevalier} \& {Li}(1999)}]{chevalier1999:apj520:29}
{Chevalier}, R.~A., {Li}, Z., 1999.
\newblock \apjl, 520, 29

\bibitem[{{Dickey} \& {Lockman}(1990)}]{dickey1990:araa28:215}
{Dickey}, J.~M., {Lockman}, F.~J., 1990.
\newblock \araa, 28, 215

\bibitem[{{Ford} et~al.(1995)}]{ford1995:apj439:307}
{Ford}, L.~A., et~al., 1995.
\newblock \apj, 439, 307

\bibitem[{{Frontera} et~al.(2000)}]{frontera2000:apjs127:59}
{Frontera}, F., et~al., 2000.
\newblock \apjs, 127, 59

\bibitem[{{Frontera} et~al.(2004)}]{frontera2004:apj614:301}
{Frontera}, F., et~al., 2004.
\newblock \apj, 614, 301

\bibitem[{{Fukugita} et~al.(1995){Fukugita}, {Shimasaku} \&
  {Ichikawa}}]{fukugita1995:pasp107}
{Fukugita}, M., {Shimasaku}, K., {Ichikawa}, T., 1995.
\newblock \pasp, 107, 945

\bibitem[{{Gehrels} et~al.(2004)}]{gehrels2004:apj611:1005}
{Gehrels}, N., et~al., 2004.
\newblock \apj, 611, 1005

\bibitem[{{Goad} et~al.(2006)}]{goad2006:aa449:89}
{Goad}, M.~R., et~al., 2006.
\newblock \aap, 449, 89

\bibitem[{{Guidorzi} et~al.(2006)}]{guidorzi2006:pasp188:288}
{Guidorzi}, C., et~al., 2006.
\newblock \pasp, 118, 288

\bibitem[{{in't Zand} et~al.(2001)}]{intzand2001:apj559:710}
{in't Zand}, J.~J.~M., et~al., 2001.
\newblock \apj, 559, 710

\bibitem[{{King} et~al.(2005){King}, {O'Brien}, {Goad}, {Osborne}, {Olsson} \&
  {Page}}]{king2005:apj630:113}
{King}, A., {O'Brien}, P.~T., {Goad}, M.~R., {Osborne}, J., {Olsson}, E.,
  {Page}, K., 2005.
\newblock \apjl, 630, L113

\bibitem[{{Kumar} \& {Panaitescu}(2000)}]{kumar2000:apj541}
{Kumar}, P., {Panaitescu}, A., 2000.
\newblock \apjl, 541, L51

\bibitem[{{Lazzati} \& {Perna}(2002)}]{lazzati2002:mnras330:383}
{Lazzati}, D., {Perna}, R., 2002.
\newblock \mnras, 330, 383

\bibitem[{{Levan} et~al.(2006)}]{levan2006:apj647:471}
{Levan}, A., et~al., 2006.
\newblock \apj, 647, 471

\bibitem[{{Martin} et~al.(1989){Martin}, {Maurice} \&
  {Lequeux}}]{martin1989:aa215:219}
{Martin}, N., {Maurice}, E., {Lequeux}, J., 1989.
\newblock \aap, 215, 219

\bibitem[{{M{\'{e}}sz{\'{a}}ros} \& {Rees}(1997)}]{meszaros1997:apj476:232}
{M{\'{e}}sz{\'{a}}ros}, P., {Rees}, M.~J., 1997.
\newblock \apj, 476, 232

\bibitem[{{Nousek} et~al.(2006)}]{nousek2006:apj642:389}
{Nousek}, J.~A., et~al., 2006.
\newblock \apj, 642, 389

\bibitem[{{O'Brien} et~al.(2006)}]{obrien2006:apj647:1213}
{O'Brien}, P.~T., et~al., 2006.
\newblock \apj, 647, 1213

\bibitem[{{Piro} et~al.(2005)}]{piro2005:apj623:314}
{Piro}, L., et~al., 2005.
\newblock \apj, 623, 314

\bibitem[{{Roming} et~al.(2005)}]{roming2005:ssrv120:95}
{Roming}, P.~W.~A., et~al., 2005.
\newblock Space Science Reviews, 120, 95

\bibitem[{{Rossi} et~al.(2002){Rossi}, {Lazzati} \&
  {Rees}}]{rossi2002:mnras332:945}
{Rossi}, E., {Lazzati}, D., {Rees}, M.~J., 2002.
\newblock \mnras, 332, 945

\bibitem[{{Sari} et~al.(1999){Sari}, {Piran} \& {Halpern}}]{sari1999:apj519:17}
{Sari}, R., {Piran}, T., {Halpern}, J.~P., 1999.
\newblock \apjl, 519, 17

\bibitem[{{Schlegel} et~al.(1998){Schlegel}, {Finkbeiner} \&
  {Davis}}]{schlegel1998:apj500}
{Schlegel}, D.~J., {Finkbeiner}, D.~P., {Davis}, M., 1998.
\newblock \apj, 500, 525

\bibitem[{{Skrutskie} et~al.(2006)}]{skrutskie2006:aj131:1163}
{Skrutskie}, M.~F., et~al., 2006.
\newblock \aj, 131, 1163

\bibitem[{{Starling} et~al.(2005)}]{starling2005:aa442:21}
{Starling}, R.~L.~C., et~al., 2005.
\newblock \aap, 442, L21

\bibitem[{{Tagliaferri} et~al.(2005)}]{tagliaferri2005:nature436}
{Tagliaferri}, G., et~al., 2005.
\newblock \nat, 436, 985

\bibitem[{{Tanvir} et~al.(2005)}]{tanvir2005:gcn3632}
{Tanvir}, N., et~al., 2005.
\newblock GCN Circular, 3632, 1

\bibitem[{{Tokunaga} \& {Vacca}(2005)}]{tokunaga2005:pasp117}
{Tokunaga}, A.~T., {Vacca}, W.~D., 2005.
\newblock \pasp, 117, 421

\bibitem[{{Wilms} et~al.(2000){Wilms}, {Allen} \&
  {McCray}}]{wilms2000:apj542:914}
{Wilms}, J., {Allen}, A., {McCray}, R., 2000.
\newblock \apj, 542, 914

\bibitem[{{Zhang} et~al.(2006){Zhang}, {Fan}, {Dyks}, {Kobayashi},
  {M{\'e}sz{\'a}ros}, {Burrows}, {Nousek} \& {Gehrels}}]{zhang2006:apj642:354}
{Zhang}, B., {Fan}, Y.~Z., {Dyks}, J., {Kobayashi}, S., {M{\'e}sz{\'a}ros}, P.,
  {Burrows}, D.~N., {Nousek}, J.~A., {Gehrels}, N., 2006.
\newblock \apj, 642, 354

\bibitem[{{Zhang} \& {M{\'e}sz{\'a}ros}(2004)}]{zhang2004:ijmp19:2385}
{Zhang}, B., {M{\'e}sz{\'a}ros}, P., 2004.
\newblock International Journal of Modern Physics A, 19, 2385

\end{thebibliography}

\label{lastpage}

\end{document}